\RequirePackage[2020-02-02]{latexrelease}
\documentclass[
aps,%
12pt,%
final,%
notitlepage,%
oneside,%
onecolumn,%
nobibnotes,%
nofootinbib,%
superscriptaddress,%
noshowpacs]%
{revtex4}
\usepackage{graphicx}
\usepackage{color}

\def\be{\begin{equation}}
\def\ee{\end{equation}}
\def\bea{\begin{eqnarray}}
\def\eea{\end{eqnarray}}
\begin{document}

\noindent {\it Astronomy Reports, 2023, Vol. , No. }
\bigskip\bigskip  \hrule\smallskip\hrule
\vspace{35mm}


\title{EXTENDED BLACK HOLE GEOMETROTHERMODYNAMICS \footnote{Paper presented at the Fifth Zeldovich meeting, an international conference in honor of Ya. B. Zeldovich held in Yerevan, Armenia on June 12--16, 2023. Published by the recommendation of the special editors: R. Ruffini, N. Sahakyan and G. V. Vereshchagin.}}

\author{\bf \copyright $\:$  2023.
\quad \firstname{Hernando}~\surname{Quevedo}}%
\email{quevedo@nucleares.unam.mx}
\affiliation{Institute of Nuclear Sciences, National Autonomous University of Mexico, AP 70543,  Mexico City 04510, Mexico}%

\begin{abstract}

\centerline{\footnotesize Received: ;$\;$
Revised: ;$\;$ Accepted: .}\bigskip\bigskip\bigskip

Although ordinary laboratory thermodynamic systems are known to be homogeneous systems, black holes are different and cannot be considered within this class. Using the formalism of geometrothermodynamics, we show that black holes should be considered as quasi-homogeneous systems. As a consequence, we argue that coupling constants in generalized gravity theories should be considered as thermodynamic variables, giving raise to extended versions of black hole thermodynamics.

\end{abstract}

\maketitle

\section{Introduction}

In classical thermodynamics, a system with $n$ thermodynamic degrees of freedom is described by means of $n$ extensive variables $E^a$, $a=1,2,..., n$, $n$ intensive variables $I_a$, and a thermodynamic potential $\Phi$ \cite{callen}. All the properties of the system can be derived from the fundamental equation $\Phi=\Phi(E^a)$, which relates all the extensive variables of the system and is demanded to satisfy, in particular, the first law of thermodynamics, which using the above notations can be written as
\be 
d\Phi = \sum_a I_a d E^a \ , \quad I_a = \frac{\partial \Phi}{\partial E^a}\ .
\label{flaw}
\ee 
In addition, the fundamental equation $\Phi=\Phi(E^a)$ should satisfy the second law of thermodynamics. In the case of ordinary laboratory systems like the ideal gas, van der Waals systems, etc., 
the function $\Phi(E^a)$ turns out to be a homogeneous function of first degree, i.e., it satisfies the relationship
\be 
\Phi(\lambda E^a) = \lambda \Phi(E^a)
\label{hom}
\ee
for any positive real value of $\lambda$, which is a consequence of the extensivity property of the variables $E^a$. A generalization of the relationship (\ref{hom}) is used to define quasi-homogeneous functions that satisfy the condition 
\be
\Phi (\lambda^{\beta_1} E^1, \lambda^{\beta_2} E^2, ..., \lambda^{\beta_n} E^n) = \lambda^{\beta_\Phi} \Phi(E^1, E^2, ... , E^n)\ ,
\label{qhom}
\ee 
where $\beta_a$ are real constants, which we call coefficients of quasi-homogeneity.

Black holes are non-ordinary thermodynamic systems. In this case, the fundamental equation is represented by the Bekenstein-Hawking entropy relation \cite{bek72,haw75} (we use geometric units with $c=G=\hbar=k_B=1)$
\be 
S = \frac{1}{4} A_h ,
\label{bhent}
\ee
where $A_h$ is the area of the black hole horizon, which according to the no-hair theorem of general relativity depends on the mass $M$ and angular momentum $J$ only \cite{heusler}. This means that the fundamental equation is given in terms of the function $S(M,J)$. The fact that the entropy of the black hole is proportional to the area and not to the volume, as in ordinary thermodynamic systems, is the first indication that black holes cannot be considered as common thermodynamic systems. In fact, one of the main conceptual problems of the thermodynamics, which follows from the Bekenstein-Hawking  fundamental equation (\ref{bhent}), is that at present there is no physically meaningful microscopic model based on a consistent definition of microstates from which black hole thermodynamics emerges as a well-defined limit. Nevertheless, black hole thermodynamics is currently a field of very active research. 

In this work, we will analyze the properties of the fundamental equation (\ref{bhent}) from a different point of view and will show that it should be interpreted as describing a quasi-homogeneous thermodynamic system, whose properties are different from those of ordinary systems. In fact, we will use the formalism of geometrothermodynamics (GTD)
\cite{quev07}
to show that, in general, fundamental equations should satisfy the quasi-homogeneity condition (\ref{qhom}) so that ordinary systems arise as a particular case when all the quasi-homogeneity coefficients are equal to one. As a consequence of imposing quasi-homogeneity, we will see that, in generalized theories of gravity, coupling constants such as the cosmological constant must be treated as thermodynamic variables, giving raise to extended versions of black hole thermodynamics.

This work is organized as follows. 
In Sec. \ref{sec:gtd}, we present a review of GTD as a formalism that uses concepts of differential geometry to analyze the properties of thermodynamic systems.
In Sec. \ref{sec:qhom}, we show that imposing quasi-homogeneity in GTD allows us to investigate thermodynamic systems in a unified manner, leading to the interesting conclusion that the geometric properties of the equilibrium space of a thermodynamic system can be used to explore its phase transition structure and stability properties. In Sec. \ref{sec:exttd}, we show that quasi-homogeneity leads to the conclusion that in generalized theories of gravity the coupling constants should be interpreted as thermodynamic variables and, consequently, the laws of thermodynamics must be generalize respectively, resulting in  extended versions of thermodynamics of black holes. Finally, in Sec. \ref{sec:con}, we summarize and comment on our results.


\section{Review of geometrothermodynamics}
\label{sec:gtd}

Let us introduce the $n-$dimensional equilibrium space ${\cal E}$ with coordinates $E^a$ and metric $g_{ab}$. 
On ${\cal E}$, it is assumed that the fundamental equation $\Phi = \Phi (E^a)$ is satisfied so that ${\cal E}$ corresponds to a specific thermodynamic system. This implies that the metric $g_{ab}$ should also be determined in terms of the function $\Phi(E^a)$ only. The idea of GTD consists in associating the geometric properties of ${\cal E}$ with the thermodynamic properties of the system. In particular, the thermodynamic interaction of the system should be associated with the curvature  and quasi-static processes should be related to the geodesics of ${\cal E}$.

One important property of classical thermodynamics is that it does not depend on the choice of thermodynamic potential. Indeed, the Legendre transformation
\be 
\Phi \rightarrow \tilde \Phi = \Phi - \sum_k E^k \frac{\partial \Phi}{\partial E^k}\ ,
\label{leg0}
\ee 
where the index $k$ takes any value from $1$ to $n$, introduces new potentials $\tilde \Phi$, which can be used to described the same system, without changing its physical properties. Since the above transformation involves $\Phi$ and its derivatives, it cannot be considered as a coordinate transformation on ${\cal E}$. To fix this problem, we introduce an auxiliary $2n+1$ dimensional phase space ${\cal T}$ with coordinates $Z^A = \{ \Phi, E^a, I_a\}$ and consider the coordinate transformation 
 of the form 
$Z^A\rightarrow \tilde Z ^A = (\tilde \Phi, \tilde E^a, \tilde I_a)$ such that 
\cite{arnold}
\be
\Phi = \tilde \Phi - \sum_k   \tilde E ^k \tilde I_k \ ,\quad
E^i = - \tilde I_{i}, \ \
E^j = \tilde E ^j,\quad
I_{i} = \tilde E ^ i , \ \
I_j = \tilde I _j \ ,
\label{leg}
\ee 
where $i\cup j$ is any disjoint decomposition of the set of
indices $\{1,...,n\}$, and $k= 1,...,i$. In particular, for
$i=\emptyset$ we obtain the identity transformation, and 
for
$i=\{1,...,n\}$, Eq.(\ref{leg}) defines a total Legendre
transformation. Furthermore, if we assume that ${\cal E}$ is a subspace of ${\cal T}$ defined by means of the smooth embedding map
\be 
\varphi: {\cal E} \rightarrow {\cal T}
\ee
or in coordinates
\be 
\varphi: \{E^a\} \rightarrow \{\Phi(E^b), E^a, I_a(E^b)\}\ ,
\ee
then one can show that on ${\cal E}$, the Legendre transformation (\ref{leg}) reduces to (\ref{leg0}). So, we see that the objective of introducing the phase space ${\cal T}$ is to be able to represent Legendre transformations as coordinate transformations. Notice that the embedding map $\varphi$ implies the fundamental equation $\Phi = \Phi(E^a)$ and, consequently, 
\be
d\Phi = \sum_a \frac{\partial \Phi}{\partial E^a} d E^a \ ,
\ee
which is essentially the first law of thermodynamics (\ref{flaw}). 

In addition, we can introduce a metric $G_{AB}$ on ${\cal T}$, which generates in a canonical way the metric $g_{ab}$ of ${\cal E}$ by means of the pullback 
\be
\varphi^*(G_{AB}dZ^A dZ^B) = g_{ab} dE^adE^b,
\ee
which implies that  
\be
g_{ab} = G_{AB} \frac{\partial Z^A}{\partial E^a} \frac{\partial Z^B}{\partial E^b}\ .
\label{pullback}
\ee
Then, we say that $g_{ab}$ is Legendre invariant if the corresponding generating metric $G_{AB}$ does not change its explicit form under the action of Legendre transformations, as expressed in Eq. (\ref{leg}). It is then possible to write down the set of algebraic equations that follow from demanding that the components of the metric $G_{AB}$ be functionally invariant with respect to Legendre transformations. The resulting system of algebraic equations can be solved analytically, leading to three different classes of solutions whose line elements can be represented as (we assume the sum convention over repeated indices)
\be
G^{^{I}}=  (d\Phi - I_a d E^a)^2 + (\xi_{ab} E^a I^b) (\delta_{cd} dE^c dI^d) \ ,
\label{GI}
\ee
\be 
G^{^{II}}= (d\Phi - I_a d E^a)^2 + (\xi_{ab} E^a I^b) (\eta_{cd} dE^c dI^d) \ ,
\label{GII}
\ee
\be	
\label{GIII}
G^{{III}}  =(d\Phi - I_a d E^a)^2  + \sum_{a=1}^n \xi_a (E_a I_a)^{2k+1}   d E^a   d I^a \ ,
\ee
where $\eta_{ab}= {\rm diag}(-1,1,\cdots,1)$, $\xi_a$ are real constants, $\xi_{ab}$ is a diagonal $n\times n$ real matrix, and $k$ is an integer. As we can see, the condition of Legendre invariance does not fix completely the form of the metric components $G_{AB}$ but leaves the coefficients $k$, $\xi_a$, and $\xi_{ab}$ arbitrary.

\section{Quasi-homogeneity}
\label{sec:qhom}

If we demand that the above metrics can be applied simultaneously to the same thermodynamic system (i.e., for a given function $\Phi(E^a)$) and lead to consistent results, one can show that the free parameters should be chosen in terms of the coefficients of quasi-homogeneity as 
\be 
\xi_{ab}={\rm diag}( \beta_1, \beta_2, ..., \beta_n) ,
\ee
\be
\xi_a = \beta_a\ .
\ee
Then, using the general expression (\ref{pullback}), from Eqs.(\ref{GI}), (\ref{GII}), and (\ref{GIII}) we obtain
\be
g^{{I}}_{ab} =   \beta_\Phi \Phi  \delta_a^{\ c}
\frac{\partial^2\Phi}{\partial E^b \partial E^c}   ,
\label{gdownI}
\ee
\be
g^{II}_{ab} =   \beta_\Phi \Phi  \eta_a^{\ c}
\frac{\partial^2\Phi}{\partial E^b \partial E^c}   ,
\label{gdownII}
\ee
\be
g^{{III}} = \sum_{a=1} ^n \beta_a \left(\delta_{ad} E^d \frac{\partial\Phi}{\partial E^a}\right)^{2k+1}
 \delta^{ab} \frac{\partial ^2 \Phi}{\partial E^b \partial E^c}
dE^a dE^c \ ,
\label{gdownIII}
\ee
respectively, 
where $\delta_a^{\ c}={\rm diag}(1,\cdots,1)$, $\eta_a^{\ c}={\rm diag}(-1,1,\cdots,1)$. 
To obtain the components of the metrics $g^I$ and $g^{II}$, we have used the  quasi-homogeneous Euler identity in the form \cite{qqs17}
 \be
 \sum_a
 \beta_a I_aE^a = \beta_\Phi \Phi\ ,  \label{egd}
\ee
which generates the conformal factor $\beta_\Phi \Phi$ of $g^I$ and $g^{II}$.
As we can see, the final expressions for the metric components $g^I_{ab}$ and $g^{II}_{ab}$ do not contain the quasi-homogeneity coefficients $\beta_a$ explicitly so that they can be  applied indistinctly to homogeneous and quasi-homogeneous systems. Notice that the conformal constant $\beta_\Phi$ does not affect the geometric properties of the corresponding Riemannian manifolds.  

To be more specific, consider the case of a system with two thermodynamic degrees of freedom ($n=2)$, i.e., the fundamental equation is determined by the function $\Phi(E^1, E^2)$. Then, 
from Eqs.(\ref{gdownI})-(\ref{gdownIII}), we obtain 
\bea
\label{gI2D} 
g^{I}& = & \beta_\Phi  \Phi \left[\Phi_{,11} (d E^1)^2 + 2 \Phi_{,12} dE^1 dE^2 + \Phi_{,22} (dE^2)^2\right]\, \\
g^{II} & = & \beta_\Phi \Phi \left[-\Phi_{,11} (d E^1)^2  + \Phi_{,22} (dE^2)^2\right]\,,
\label{gII2D} \\
g^{III} & = & \beta_1 (E^1 \Phi_{,1})^{2k+1} \Phi_{,11} (dE^1)^2 
+ \beta_ 2 (E^2 \Phi_{,2})^{2k+1} \Phi_{,22} (dE^2)^2 \nonumber \\
& & + \left[\beta_1 (E^1\Phi_{,1})^{2k+1}  + \beta_2 (E^2\Phi_{,2})^{2k+1} \right]\Phi_{,12} dE^1 dE^2
\ ,
\label{gIII2D}
\eea
where $\phi_{,a} = \frac{\partial \phi}{\partial E^a}$, etc.

We demand again that all the line elements $g^I$, $g^{II}$, and $g^{III}$ can be used to describe the same system. This means that in particular the curvature of the three metrics should lead to compatible results. It is then easy to show that this condition implies that $k=0$, which should also be valid for arbitrary systems with arbitrary number of degrees of freedom. 

The computation of the curvature scalars results in  
\be
R^I = \frac{N^I}{D^I}\ ,\ \ D^I = 2 \beta_\Phi \Phi ^3 
\left[\Phi_{,11}\Phi_{,22} -(\Phi_{,12})^2
\right]^2\ ,
\label{denoi}
\ee
\be
R^{II} = \frac{N^{II}}{D^{II}}\ ,\ \ D^{II}  = 2\beta_\Phi \Phi ^3
\left( \Phi_{,11} \Phi_{,22}\right)^2\ ,
\label{denoii}
\ee
\be
R^{III} = \frac{N^{III}}{D^{III}}\ ,\ \ D^{III}  = 
\left[ \beta_\Phi ^2 \Phi^2 (\Phi_{,12})^2
- 4 \beta_1 \beta_2 E^1   E^2  \Phi_{,1} \Phi_{,2} \Phi_{,11} \Phi_{,22}
\right]^3 \ ,
\label{denoiii}
\ee
respectively, where  we have used the Euler identity in the form
\be
\beta_1 E^1 \Phi_{,1} + \beta_2 E^2 \Phi_{,2} = \beta_\Phi \Phi\ ,
\ee
to simplify the final form of the function $D^{III}$. The functions
$N^I$, $N^{II}$ and $N^{III}$ depend on $ \Phi $ and its derivatives. In general, they can be shown to be non-zero when the denominators $D^I$, $D^{II}$, and $D^{III}$ vanish, which determine the locations of curvature singularities. 

The condition  $D^I=0$ implies that $\Phi_{,11} \Phi_{,22} = (\Phi_{,12})^2$ so that and $D^{II} \neq 0$ and
\be
D^{III}  = (\Phi_{,12})^6
\left[ \beta_\Phi ^2 \Phi^2 
- 4 \beta_1 \beta_2 E^1   E^2  \Phi_{,1} \Phi_{,2} \right]^3 \ .
\ee
Then, one can see that the expression inside the parenthesis is zero only if $\Phi$ depends on one variable only. 
A further detailed analysis shows that 
 all the singularities are determined by the zeros of the second-order derivatives of $\Phi$, namely, 
\bea
I: &&  \Phi_{,11}\Phi_{,22} -(\Phi_{,12})^2
=0 \ ,\label{singirev} \\
II: &&  \Phi_{,11} \Phi_{,22} 
=0\ , \label{singiirev}  \\
III: &&  \Phi_{,12}= 0 \ . \label{singiiirev} 
\eea
This shows that the singularity structure of all the above metrics is given in terms of three compatible conditions. This is the result of imposing quasi-homogeneity at the level of the corresponding metrics.
Interestingly, the above equations are exactly the conditions that determine the stability properties of thermodynamic systems \cite{callen}.


\section{Extended thermodynamics}
\label{sec:exttd}

In the last section, we showed that the condition of quasi-homogeneity allows us to fix the arbitrary parameters of the GTD metrics in such a way that they can be applied in a consistent manner to any thermodynamic system, leading to compatible results at the level of the curvature singularities.

In this section, we will explore the consequences of imposing quasi-homogeneity at the level of the fundamental equations of black holes. Consider the Bekenstein-Hawking entropy of the most general black hole in Einstein's gravity theory 
\cite{smarr73,davies78}
\be
S= 2\pi \left( M^2+\sqrt{M^4-J^2}\right)\ ,
\ee
where $M$ is the mass and $J$ the angular momentum of the Kerr black hole. The quasi-homogeneity condition
\be
S(\lambda^{\beta_M} M , \lambda^{\beta_J} J) =
2\pi \lambda^{2\beta_M} \left(M^2
+\sqrt{M^4 - \lambda^{2\beta_J-4\beta_M} J^2} \right)
\ee
is satisfied if the relationship
\be
\beta_J = 2 \beta_M\ , \ \ \beta_S = 2 \beta_M 
\ee
is fulfilled. This implies that, in fact, the Kerr black hole can be considered as a quasi-homogeneous system of degree $\beta_S = 2 \beta_M$.

Consider now Einstein theory with cosmological constant $\Lambda$, which is described by the action 
\be
{\cal S} = \frac{1}{16\pi} \int d^4 x \,\sqrt{-g} \left( R - 2\Lambda \right) \ .
\ee
In this case, the most general solution representing a black hole configuration is known as the Kerr-AdS solution \cite{solutions}, whose fundamental equation can be expressed as \cite{cck00}
\be
M^2=J^2 \left(-\frac{\Lambda}{3}+ \frac{\pi}{S}\right)
+ \frac{S^3}{4\pi^3}\left(-\frac{\Lambda}{3}+ \frac{\pi}{S} 
\right)^2 \ .
\label{feknads4}
\ee
If we perform the rescaling  
$M\rightarrow \lambda^{\beta_M} M$, 
$S\rightarrow \lambda^{\beta_S} S$, 
$J\rightarrow \lambda^{\beta_J} J$,
we can see that the function (\ref{feknads4}) does not satisfy  the quasi-homogeneity 
condition. However, if we consider the cosmological constant $\Lambda$ as a thermodynamic variable which rescales as 
$\Lambda\rightarrow \lambda^{\beta_\Lambda} \Lambda$, 
the fundamental equation (\ref{feknads4}) is a quasi-homogeneous function if  
\be
\beta_J = \beta_S\ , \ \beta_\Lambda = - \beta_S \ , \ 
\beta_M = \frac{1}{2}\beta_S\ .
\ee 
This means that thermodynamic description of the Kerr-AdS black hole must extended to include the additional variable $\Lambda$. Indeed, it has been shown 
\cite{krt09,cgkp11,kmt17}
that the cosmological constant can be interpreted as an effective pressure so that, in particular, the first law of thermodynamics should be written as
\be
dM = T d S + \Omega d J + V d\Lambda \ ,
\ee
where $T$ is the temperature, $\Omega$ the angular velocity at the horizon, and $V$ is the effective volume of the black hole. This extended version of black hole thermodynamics has been proved to lead to a very rich structure of the phase transition structure of the Kerr-AdS black hole, a structure that resembles the behavior of van der Waals systems. For this reason this extended version of thermodynamics is known as black hole chemistry \cite{kmt17}.

It can be shown that in other generalized gravity theories, such as Bord-Infeld, Lovelock, Yang-Mills, etc., the coupling constants that appear in the corresponding black hole solutions must be considered as thermodynamic variables as a consequence of the quasi-homogeneity condition. This opens the possibility of extending black hole thermodynamics by including new variables, which is expected to lead to new and interesting aspects of black hole physics.

\section{Conclusions}
\label{sec:con}

In this work, we have shown that by demanding compatibility within the  formalism of GTD, it follows that thermodynamic systems can be split into ordinary homogeneous and quasi-homogeneous systems. As a particular example of the last class, we mention black holes in Einstein theory and its generalizations. 

Moreover, we have shown that the property of quasi-homogeneity implies that the coupling constants, which  appear in generalized gravity theories, must be considered as thermodynamic variables. 
In particular, in the case of AdS black holes, the cosmological constant can be treated as an effective pressure, leading to a generalization of the corresponding laws of black hole thermodynamics, 
an approach that is known in the literature as extended black hole thermodynamics. 

Thus, we conclude that the geometrothermodynamic approach, which we use to explore the properties of thermodynamic systems, in general,  can also be applied to find new aspects of the thermodynamic structure and the physics of black holes.

\begin{acknowledgments}

I thank the organizers of the 5th Zeldovich Meeting for the excellent organization and hospitality during the development of the Meeting in Yerevan.
\end{acknowledgments}

\section*{Funding}
This work was partially supported  by UNAM-DGAPA-PAPIIT, Grant No. 114520, Conacyt-Mexico, Grant No. A1-S-31269.

\clearpage


\end{document}